\documentclass[aps,prl,a4paper,twocolumn,amsmath,amssymb,floatfix,
superscriptaddress,showpacs]{revtex4}
\usepackage[dvips]{graphicx}

\begin{document}

\title{Griffiths phase in the thermal quantum Hall effect}
\author{A. Mildenberger}
\affiliation{Institut f\"ur Nanotechnologie, Forschungszentrum Karlsruhe,
76021  Karlsruhe, Germany}

 \author{F. Evers}
\affiliation{Institut f\"ur Nanotechnologie, Forschungszentrum Karlsruhe,
76021  Karlsruhe, Germany}

\author{R. Narayanan} 
\affiliation{Institut f\"ur Nanotechnologie, Forschungszentrum Karlsruhe,
76021  Karlsruhe, Germany}

\author{A.~D. Mirlin}
\altaffiliation[Also at ]{Petersburg Nuclear Physics
Institute, 188350 St.~Petersburg, Russia.}
\affiliation{Institut f\"ur Nanotechnologie, Forschungszentrum Karlsruhe,
76021  Karlsruhe, Germany}
\affiliation{Institut f\"ur Theorie der Kondens. Materie,
Universit\"at Karlsruhe, 76128 Karlsruhe, Germany}

\author{K. Damle}
\altaffiliation[Previously at ]{Dept. of Physics and Astronomy, Rice University,
Houston, TX 770 005}
\affiliation{Dept. of Theoretical Physics,
Tata Institute of Fundamental Research,
Mumbai 400005, India}

\date{\today}

\begin{abstract}
Two dimensional disordered superconductors with broken spin-rotation and
time-reversal invariance, e.g. with $p_x+ip_y$ pairing, 
can exhibit plateaus in the thermal Hall coefficient (the thermal quantum Hall
effect).  
Our numerical simulations show that the Hall insulating regions of
the phase diagram can 
support a sub-phase where the quasiparticle density of states is divergent at
zero energy,
$\rho(E)\sim |E|^{1/z-1}$, with a non-universal exponent $z>1$, due to
the effects of rare configurations of disorder (``Griffiths phase''). 

\end{abstract}

\maketitle

The {\it integer quantum Hall effect} (IQHE)
is observed in two-dimensional (2d) electron gases
in high magnetic fields. 
Its striking manifestation is the quantization of the Hall
conductance $\sigma_{xy}$ in units of ${e^2\over h}$.
Analogs of the 
IQHE are also known in systems with broken time reversal symmetry in which
quasiparticle charge conservation is violated because of the
presence of a charge condensate, while quasiparticle spin and energy
remain conserved.
In such a situation the 
Hall coefficients for spin or heat transport
can exhibit plateaus at integer values in units
of  $\hbar/2\pi$ \cite{senthil99}
or $\pi k_B^2/3\hbar$ \cite{senthil00,vishwanath01}, respectively. 
These novel types of quantum  Hall effects may be encountered
in superconducting systems with unconventional
(non-s-wave) pairing. There has been a great deal of recent interest
in these
{\it spin and thermal quantum Hall effects} (SQHE and TQHE)
\cite{bundschuh99,gruzberg99,beamond,kagalovsky,senthil99,mirlin,senthil00,vishwanath01}. 
In particular, the TQHE~\cite{chalker01} in systems {\em without} spin conservation, which
belongs to class D in the Altland-Zirnbauer
scheme~\cite{altland}, displays many peculiar features that are not found in the IQHE or
SQHE, including a {\it metallic} phase with a logarithmically diverging
density of states (DoS) \cite{senthil00,bocquet00}. More generally, behavior
of the DoS in unconventional disordered superconductors has attracted a lot of
research interest recently \cite{hirschfeld02}.  A wealth
of different types of low-energy behavior of the
DoS was identified, depending on the symmetry of the
system and on the type of disorder.

In a recent study of the Random Bond Ising model (RBIM) (which falls in the
same universality class as the TQHE systems),
Gruzberg et al. \cite{gruzberg01} conjectured
that such systems may have Hall
insulating phases supporting a region where the DoS exhibits a power law singularity
$\rho(E)\propto |E|^{-1+1/z}$ with a nonuniversal
exponent $z$. Strong-disorder renormalization group (RG) calculations of Motrunich et al. \cite{motrunich01}
have also demonstrated the presence of such low $E$ divergence in the DoS (with $z > 1$) in
quasi-1d superconductors with broken time reversal and spin rotation invariance. This has also been confirmed by calculations using a Fokker-Planck approach~\cite{gruzberg02}.
The strong-disorder RG calculations make it apparent that the DoS divergence
originates in this and in other $1$d models~\cite{damle02}
from {\em Griffiths effects}~\cite{griffiths69}
involving exponentially rare length $\ell$ regions (occuring with
probability $\sim e^{-c'\ell}$) of the sample in `nearby' phases.
An exponentially weak coupling, $\sim e^{-c\ell}$, 
of defects at the boundaries of such rare regions produces a power-law tail in the low energy DoS:
\begin{equation}
\label{e1}
\rho(E)  \sim \int \! d\ell \ \delta \! \left( E - e^{-c\ell} \right) 
e^{-c'\ell} \propto |E|^{1/z-1},
\end{equation}
($z{=}c/c'$ and $c,c'$ are non universal parameters).
Motrunich {\it et al.}
further speculated  \cite{motrunich01,motrunich02}
that a similar mechanism driving a divergent DoS may be operative
also in 2d superconductors.

The presence of such a regime with divergent DoS is in sharp contrast to the situation
in standard Wigner-Dyson ensembles, where the DoS is never divergent. In 2d
power law divergences generically modified by logarithmic corrections have been known to
exist in the chiral symmetry classes ~\cite{gade93,ludwig94,mudry03,motrunich02}, where
they appear in critical phases without exponential localization of the wavefunctions. 
Here, we report an observation of an {\it insulating} Griffiths phase
with power-law singularities in the DoS of a model (RBIM) in the
TQHE universality class.
We further demonstrate that the divergent DoS is
governed by a mechanism involving the presence of rare long {\em string}
defects~\cite{foot1}.
Our numerical results thus show that TQHE
-- in contrast to its close relatives, the IQHE and the SQHE --
does exhibit the conjectured Griffiths phase
\cite{gruzberg01,motrunich01}. 

\begin{figure}[t]
\centering
\includegraphics[width=6.0truecm]{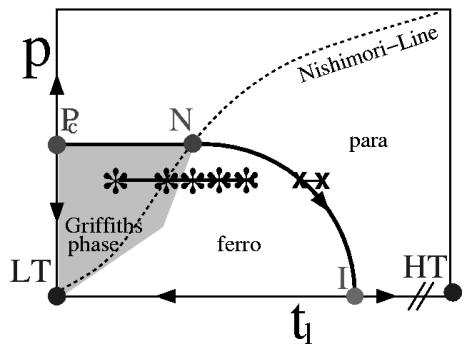}
\caption{Phase diagram of TQHE in 
  RBIM-network model representation, disorder $p$ vs. inter plaquette
  coupling strength (temperature) $t_l$. Fixed points: high temperature (HT), low temperature (LT),
  clean Ising (I), multicritical point (N), percolation (${\rm P}_{\rm
  c}$). Phase boundary (solid) separates two Hall insulators
  (called ferro and para using RBIM terminology). Shaded
  region indicates Griffiths phase with diverging
  single particle density of states, $\rho(E)$. Line segments represent
  scans for $\rho(E)$ displayed in Figs. \ref{f2} (x)
  and \ref{f3} (*). In Fig. \ref{f4} scan is along phase
  boundary from $N$ to $P_c$.}
\label{f1}
\vspace*{-0.30cm} 
\end{figure}

For our numerical simulations we choose a particular
network realization of the $\pm J$ RBIM. 
As we discuss in the end of the paper, our results are however not restricted
to this model but rather are generally valid for TQHE systems. 
The RBIM network realization 
is defined on a square lattice and has been described
in detail before \cite{chalker01}.
Particles traversing the links of the network change
direction and turn left or right with probability amplitudes
$\pm \cos \alpha$ or $\pm \sin\alpha$
whenever arriving at a node. 
(The respective signs are fixed by the condition of unitarity
of the corresponding time evolution operator.) 
Disorder in the form of ``vortex pairs'' 
is introduced by flipping the sign of both
$\cos\alpha$
amplitudes
of the node that is dressed by the vortex pair. 
The system is characterized by two parameters: the inter-plaquette coupling
strength, $t_l=\sin^2\alpha$ related to the temperature $T$ and coupling
constant $J$ of the Ising model as  
$t_l=\cosh^{-2} 2J/T$, and the concentration
of defects, $p$. Note that although the inter-plaquette hopping
amplitudes are very different from the intra-plaquette amplitudes,
the underlying spin model has no explicit dimerization in its couplings.
Fig.~\ref{f1} shows the phase diagram of the model. 
It contains two localized phases corresponding to the ferro- and paramagnetic
phases of the RBIM. The precise location of the phase transition line
(the thermal quantum Hall transition in the fermionic formulation)
has been studied in Ref.~\cite{merz01}. We also show in Fig.~\ref{f1} the
Nishimori line (NL), where the RBIM possesses some additional symmetry.

We construct the time evolution operator $U$
for a single time step for networks with periodic boundary
conditions and use standard sparse matrix packages to extract
a number of eigenvalues closest to zero
\cite{arpack}. Typically 8 to 16 eigenvalues per system
are calculated.
Since $U$ is unitary and the network is class $D$, the eigenvalues
come in pairs and are of the form $\exp({\pm i E})$. We obtain the
density of (pseudo)-energies, $\rho(E)$, from an average over
an ensemble of disorder realizations which contains typically
of the order of $10^4$ members. 

\begin{figure}[t]
\centering
\includegraphics[width=7.0cm]{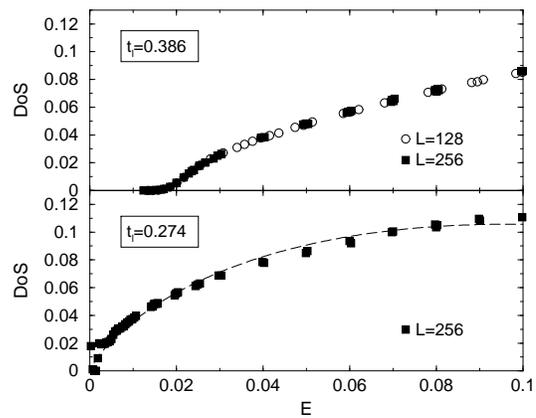}
\caption{DoS near (top) and on (bottom) of phase boundary at
  $p{=}0.08$ (see Fig. \ref{f1}).
  Top: Bare gap with exponential smearing by Lifshitz mechanism.
  Numerical fit of gap region is compatible with 
  $\ln \rho {\propto} E^{-\kappa}$, $\kappa\approx 2$.
  Bottom: Pseudo-gap in DoS fitted to the analytical prediction 
$\rho(E){\approx} (g_{\rm M}/2\pi^2)  |E|\ln |E_1/E|$; fit: $g_{\rm M}{\approx} 21.51$ and $E_1{=}0.262$. 
}
\vspace*{-0.5cm} 
\label{f2}
\end{figure}

We start the presentation of our numerical results with a scan along
the line $p=0.08$, as shown in Fig. \ref{f1}.
The scan begins in the high-$T$ (paramagnetic) phase 
where the impact of a small concentration of vortex disorder 
is relatively weak.
Here, one expects the energy gap that exists in the clean limit
($p{=}0$) to dominate because
the particles  have an appreciable tendency to wind about those
plaquettes of the sublattice that are defect free.
A non-zero DoS near zero energy is
only due to very rare configurations of vortices giving
rise to  a ``Lifshitz tail '' \cite{liftshitz},
$\ln \rho(E) \sim E^{-2} $, see top panel of Fig.~\ref{f2}
that turn the true gap of the clean system  into a ``pseudo-gap''.

At lower values of the interplaquette coupling, $t_l\approx 0.29$,
{\it i.e.} right on the phase boundary,
the shape of the DoS is more complicated. In the absence of disorder, the DoS
is linear, $\rho(E)\propto |E|$, at the Ising critical point. In a disordered
system, the theory of Dirac fermions with a random mass predicts  
a logarithmic correction to the low-energy behavior,
$\rho(E) \propto |E|\ln |E|^{-1}$ \cite{bocquet00}. Indeed our data for not
too high $E$ are consistent with this form, see bottom panel of
Fig.~\ref{f2} (at higher energies model-specific band structure effects become
important). Small deviations from this formula at very low energies can be
attributed to finite-size effects, which are particularly pronounced at
criticality.

\begin{figure}[t]
\centering
\includegraphics[width=7.0cm]{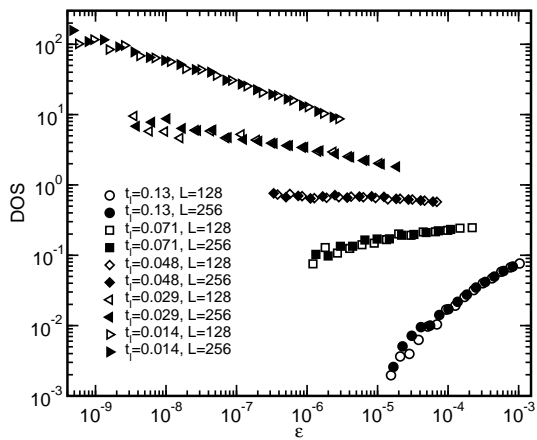}
\caption{DoS at disorder value $p=0.08$ deep in the ferromagnetic phase.
As coupling between plaquettes is progressively reduced
pseudo-gap gives way to power law divergency.
Plots show averages for two system sizes, $128$ (256)
with 20000 (10000) realizations.
Data comprises 8 lowest lying eigenstates;
$t_l=0.029$ corresponds to location on 
Nishimori line.}
\label{f3}
\vspace*{-0.35cm} 
\end{figure}

If we diminish the inter-plaquette coupling strength so as to get
close to the NL (still keeping $p=0.08$ constant),
we witness a qualitative change in the low-energy DoS:
the pseudo-gap vanishes and the DoS starts to increase with lowering $E$,
apparently diverging as $E\to 0$. This is demonstrated in
Fig. \ref{f3}, where the evolution of DoS is shown as $t_l$
is lowered from $0.13$ down to $0.014$.
For large enough inter-plaquette 
coupling one still notices the pseudo-gap behavior.
The latter is already less evident in the data that
corresponds to the coupling  strength $t_l=0.07$.
It finally gives way to a strongly increasing behavior
once the coupling falls below $t_l\approx 0.05$.

Having demonstrated the divergence of the DoS in the low-temperature part of
the ferromagnetic phase, we turn to a discussion of the corresponding
mechanism. We will show that the DoS singularity is governed by rare disorder
configurations containing long chains ({\it strings}) 
of defects. Since each defect contains
a pair of vortices, such a string of a length $\ell>1$ (formed with a
small but non-zero probability $\sim p^\ell$) 
produces two vortices with a large separation $\ell$. 
At zero plaquette coupling, $t_l=0$, every vortex contributes
a zero mode (or ``mid-gap state'') to the spectrum which
therefore has a macroscopic degeneracy $pL^2$. 
Upon switching on the coupling $t_l$, the zero modes couple 
and the degeneracy is lifted. For a small $t_l$, 
the energy splitting $E$ thus produced
will be exponentially small in $l$~\cite{read0}. For a clean system (which contains only
two vortices at the ends of the string but no other defects) this follows
immediately from the perturbation theory, yielding $E\propto 
t_l^{\ell/2}=e^{-c\ell}$. 
Clearly, the same result holds for those rare configurations
of the disordered system for which the region of a size $\sim \ell$ around the
vortex pair is free of other vortices. However, the probability for this to
happen scales with $\ell$ as $(1-p)^{\ell^d}$, where $d=2$ is the spatial
dimensionality. In the 1d case this factor is harmless, yielding the
probability for formation of an isolated string $\propto e^{-c'\ell}$ with 
$c'=-\ln p(1-p)^2$ and Eq.~\ref{e1} applies. On the 
other hand, in $d=2$ the above factor 
seems to spoil the reasoning leading to Eq.~(\ref{e1}). So, if the
requirement of ``isolation'' of the vortex pair would be a necessary condition
for the appearance of exponentially small eigenvalues, 
$\epsilon\sim e^{-c\ell}$, the divergent behavior of the DoS would be
converted to a pseudo-gap at lowest energies due to the  $(1-p)^{\ell^d}$
factor.  

\begin{figure}[t]
\centering
\includegraphics[width=7cm]{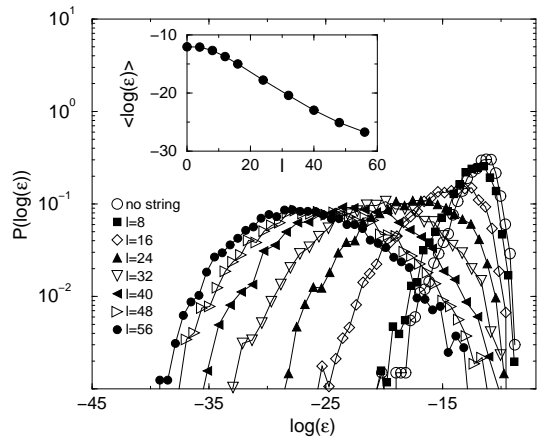}
\caption{Distribution of logarithm of lowest eigenvalue for string
  of defects of length $\ell=8\div 56$ 
inserted in disordered system in 
ferromagnetic phase ($p=0.08$, $t_l=0.029$, $L=128$). 
Inset: typical lowest eigenvalue as a function of $\ell$.}
\label{f4}
\vspace*{-0.35cm} 
\end{figure}

We now argue that this requirement is not necessary, and
Eq.~(\ref{e1}) is applicable in 2d as well.  Indeed, at small $p$ almost all
randomly located defects will be isolated. Correspondingly, $\simeq p$ 
is the concentration of
vortex pairs with separation unity and an energy splitting $\sim
t_l^{1/2}$. There will be a
much smaller concentration ($p^2$) of defects forming chains of length two,
yielding the energy splitting $\sim t_l$, and so forth. 
the energy
splitting associated with these short (compared to $\ell$) vortices will be
relatively big, and will only weakly affect the energy splitting $t^{\ell/2}$
of the $\ell$-vortex. This argument can be converted to a
RG procedure, as in \cite{motrunich01,motrunich02}. The vortices can be sorted  
into vortex pairs by combining those with the shortest distance first, 
the second shortest distance next, etc. Since the energy splitting
associated with the coupling of two vortices is exponential in their
distance, vortex pairs differing in this distance will strongly differ
in their splitting as well.  The hierarchical order allows for a
RG procedure that successively eliminates
vortex pairs with an energy splitting that is large as compared to the 
pairs remaining on the lattice. Clearly, for sufficiently small $t_l$ and $p$ 
(the boundary of the Griffiths phase will be discussed below)
the eigenvalues associated with long strings, 
i.e. vortex pairs with a large separation, 
are (largely) insensitive to the presence of other vortex pairs.  
In other words, a string  of length $\ell$ does not have to be isolated 
in order to  contribute an eigenvalue $\sim e^{-c\ell}$. The only role of
other (randomly distributed) vortices will be in somewhat modifying the
parameter $c$ as compared to its value $c=-{1\over 2}\ln t_l$ for an isolated
string.

\begin{figure}[t]
\centering
\includegraphics[width=7.0cm]{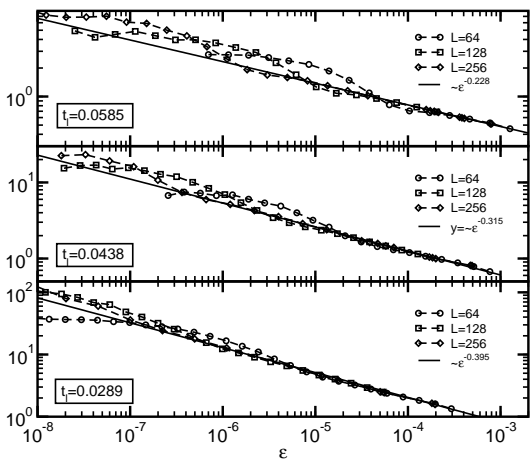}
\caption{Scan for DoS along phase boundary 
starting from Nishimori point (N) terminating at 
${\rm P}_{\rm c}$. DoS shows
power-law divergence, $\rho(E) \sim |E|^{1/z-1}$, with non-universal
exponent $z$. At low energies random-matrix oscillations indicate
existence of delocalized (critical) states.
Plots with 
$t_l=0.058,0.044,0.029$ (from top to bottom); each for
three values of system size: $L=64,128,256$.}
\label{f5}
\vspace*{-0.75cm}
\end{figure}

Since the validity of the above RG-type argument
is crucial for the proposed 
string mechanism, we have performed a numerical test for its verification. 
We have inserted a string of the length $\ell$ (ranging from 8 to 56) into a
disordered network with $t_l=0.029$ and $p=0.08$. Fig. \ref{f4} displays the
evolution of 
the distribution of the lowest eigenvalue with increasing length of the
string. The results fully confirm our scenario: 
a) the extra string gives rise to an additional, isolated
eigenvalue; b) this eigenvalue moves with increasing $l$
according to the expected exponential law, 
$\langle -\ln \epsilon\rangle\propto \ell$, see inset of Fig. \ref{f4} 
(the rest of the spectrum does not move). 
We have performed the same calculation in the paramagnetic phase ($p=0.2$,
$t_l=0.029$) and found that the distribution does {\it not} move with $\ell$: 
a long string of defects in a random environment does not generate an
exponentially small eigenvalue.

Therefore, the analytical arguments together with the numerical results, 
Figs.~\ref{f3} and \ref{f4}, demonstrate the existence of 
a Griffiths phase where string formation generates a 
divergent DoS, Eq.~(\ref{e1}) with  $z>1$.  
This phase is shaded in the phase diagram, Fig.~\ref{f1}.
Its boundary on the side of larger $t_l$
is given by the condition $z=1$ and, according to our numerical results, is
at least close to the NL. 
The asymptotic behavior of
the boundary at low temperature and weak disorder 
can be easily found from our analytical
considerations. Indeed, in this limit $c\simeq-{1\over 2}\ln t_\ell\simeq
2J/T$, 
and $c'=-\ln p$, so that the condition $z=c/c'=1$ reduces to
$\ln p \simeq -2J/T$.
This agrees with the low-temperature asymptotics of the NL, whose
exact form is $e^{2J/T}=p/(1-p)$. 
It is thus plausible that the boundary of the Griffiths phase with divergent
DoS in fact coincides with the
NL as was conjectured  in \cite{gruzberg01}.

The boundary at large disorder is set by the phase
transition line into the paramagnetic phase, where the string mechanism ceases
to be operative. Indeed, our numerics indicates pseudogap formation
everywhere in the paramagnetic phase. Right at this part of the phase 
boundary (between the Nishimori point N and the percolation critical point
${\rm P}_{\rm c}$), 
we find a divergent DoS with a non-universal exponent $z$, 
Fig.~\ref{f5}.
While we do not have at present 
an analytical theory for this behavior of DoS at criticality, 
an analogy with the random phase  XY model, which has   
a phase diagram remarkably similar to Fig.~\ref{f1}, with a low-T 
part of the phase diagram requiring a strong-disorder RG treatment, \cite{carpentier99} might be instructive.


In conclusion, we have reported the observation of a novel
type of insulating phase for 2$d$ electron systems that occurs in
the
TQHE
characterized by a divergent
density of states. Our numerical work is based on the
network representation of the $\pm J$ random bond Ising 
model. Preliminary results for the more generic 
Cho-Fisher \cite{cho97} model confirm that a similar phase
exists also there, so that the Griffiths phase
is a ubiquitous  companion of the TQHE. 

 We acknowledge valuable discussions with  
J.T.~Chalker, I.A.~Gruzberg, A.W.W.~Ludwig, C.~Mudry,       
X.~Wan, and P.~W\"olfle. This work was supported by the DFG under SPP ``Quanten-Hall-Systeme''.
K.D. was supported by a Sloan Fellowship while
at Rice University.


\vspace*{-0.80cm}



\begin{thebibliography}{99}
\bibitem{senthil99}
T. Senthil, J.B. Marsten and M.P.A. Fisher, Phys. Rev. B. 
{\bf 60}, 4245, (1999).
\bibitem{vishwanath01} A. Vishwanath, Phys. Rev. Lett. {\bf 87}, 217004 (2001).
\bibitem{senthil00} T. Senthil and M.P.A. Fisher, Phys. Rev. B {\bf
  61}, 9690 (2000).
\bibitem{bundschuh99}
  R. Bundschuh {\it et al.}, Phys. Rev. B {\bf 59}, 4382 (1999).
\bibitem{gruzberg99} I.A. Gruzberg, A.W.W. Ludwig and N. Read,
Phys. Rev. Lett. {\bf 82}, 4524 (1999).
\bibitem{beamond} E. Beamond, J. Cardy and J.T. Chalker, 
Phys. Rev. B {\bf 65}, 214301 (2002).
\bibitem{kagalovsky} V. Kagalovsky {\it et al.}, Phys. Rev. Lett. {\bf 82}, 3516 (1999).
%
\bibitem{mirlin} A.D. Mirlin, F. Evers, and A. Mildenberger,
  J. Phys. A: Math. Gen. {\bf 36}, 3255 (2003);
  F. Evers, A. Mildenberger, and  A.D. Mirlin, 
Phys. Rev. B {\bf 64}, R041303 (2003).
\bibitem{chalker01} J.~T. Chalker {\it et al.},
  Phys. Rev. B {\bf 65}, 012506 (2001).
\bibitem{altland} A. Altland, and M.R. Zirnbauer, Phys. Rev. B {\bf 55},1142
  (1997); M.R. Zirnbauer, J. Math. Phys. {\bf 37}, 4986 (1996). 
\bibitem{bocquet00}
  M. Bocquet, D. Serban, and M. R. Zirnbauer, Nucl. Phys. B {\bf 578},
  628 (2000).
\bibitem{hirschfeld02} for a review see e.g. P.J.~Hirschfeld and W.A.~Atkinson, J. Low
  Temp. Phys. {\bf 126}, 881 (2002). 
\bibitem{gruzberg01} I.A. Gruzberg, A.W.W. Ludwig and N. Read, 
Phys. Rev. B {\bf 63}, 104422 (2001).
\bibitem{motrunich01} O. Motrunich, K. Damle and D. Huse,   
Phys. Rev. B. {\bf 63}, 224204 (2001).
\bibitem{gruzberg02}  I.A. Gruzberg, N. Read, S. Vishveshwara, cond-mat/0412413.
\bibitem{damle02} K.~Damle, Phys. Rev. B {\bf 66}, 104425 (2002).  
\bibitem{griffiths69} R. B. Griffiths, Phys. Rev. Lett. {\bf 23}, 17 (1969).
\bibitem{ludwig94} A.W.W. Ludwig {\it et al.}, Phys. Rev. B {\bf 50}, 7526 (1994).

\bibitem{gade93} R. Gade, Nucl. Phys. B. {\bf 398}, 499, (1993); 
F. Wegner and R. Gade, Nucl. Phys. B. {\bf 360}, 213, (1991).
\bibitem{mudry03} C. Mudry, S. Ryu and A. Furusaki,  
Phys. Rev. B {\bf 67}, 064202 (2003).
\bibitem{motrunich02} O. Motrunich, K. Damle and D. Huse,   
Phys. Rev. B {\bf 65}, 064206 (2002).
\bibitem{foot1}There is a similarity between our results and those for the chiral symmetry class under special perturbations driving the system to the localized phase, where the power-law DoS singularity related to string formation was reported~\cite{motrunich02}.
\bibitem{read0} N.~Read and A.~W.~Ludwig, Phys. Rev. B {\bf 63}, 024404 (2001).
\bibitem{carpentier99} D. Carpentier and P. Le Doussal,
  Nucl. Phys. B {\bf 588}, 565 (2000).
\bibitem{merz01}
  F. Merz and J.T. Chalker, Phys. Rev. B {\bf } (2001).
  \bibitem{arpack}J.W. Demmel {\it et al.}, SIAM
  J. Matrix. Anal. Appl. {\bf 20}, 720 (1999); R. B. Lehoucq,
  D. Sorensen, and C. Yang, {\it ARPACK Users guide} (SIAM,
  Philadelphia, 1998). 
\bibitem{liftshitz} I.M. Lifshitz, 
Sov. Phys. JETP {\bf 17}, 1159 (1963). 
%
\bibitem{evers01} F. Evers, A. Mildenberger, A.D. Mirlin,
  Phys. Rev. B. {\bf 64}, R241303 (2001).
\bibitem{cho97} S. Cho and M.P.A.~Fisher, Phys. Rev. B {\bf 55}, 1025 (1997).

\end{thebibliography}
\end{document}